\documentclass[nofootinbib,aps,superscriptaddress,11pt]{revtex4-1}
\usepackage{geometry}
\usepackage{graphicx}
\usepackage{dcolumn}
\usepackage{bm}
\usepackage{epsfig,amsmath}
\usepackage{amssymb}
\usepackage{subfigure,esint}
\usepackage{float}
\usepackage[usenames,dvipsnames]{color}
\usepackage[pagebackref=false, colorlinks=true]{hyperref}
\definecolor{redish}{rgb}{0.7,0.2,0.0}  
\definecolor{bluish}{rgb}{0.2,0.5,0.8}

\hypersetup{linkcolor=redish,          
	citecolor=blue,        
	filecolor=magenta,      
	urlcolor=blue}          

\begin{document}
\author{Kunal Pal}\email{kunalpal@iitk.ac.in}
\affiliation{
	Department of Physics, Indian Institute of Technology Kanpur, \\ Kanpur 208016, India}
\author{Kuntal Pal}\email{kuntal@iitk.ac.in}
\affiliation{
	Department of Physics, Indian Institute of Technology Kanpur, \\ Kanpur 208016, India}
\author{Tapobrata Sarkar}\email{tapo@iitk.ac.in}
\affiliation{
	Department of Physics, Indian Institute of Technology Kanpur, \\ Kanpur 208016, India}
\title{\Large Geodesically completing regular black holes by the Simpson-Visser method}

\begin{abstract}
Regular black holes are often geodesically incomplete when their extensions to 
negative values of the radial coordinate are considered.	
Here, we propose to use the Simpson-Visser method of regularising a singular spacetime, and apply it to a regular solution that is geodesically incomplete, 
to construct a geodesically complete regular solution. Our method is generic, and can be used to cure geodesic incompleteness in any
 spherically symmetric static regular solution, so that the resulting solution is symmetric in the radial coordinate. As an example, we illustrate this 
procedure using a regular black hole solution with an asymptotic Minkowski core.
We study the structure of the resulting metric, and show that it can represent 
a wormhole or a regular black hole with a single or double horizon per side of the throat. 
Further, we construct a source Lagrangian for which the geodesically complete spacetime 
is an exact solution of the Einstein equations, and show that this consists of a phantom scalar field 
and a nonlinear electromagnetic field. Finally, gravitational lensing properties of the geodesically complete spacetime are briefly studied.
\end{abstract}

\maketitle	
\section{Introduction}
The appearance of spacetime singularities in general relativity (GR) is a generic feature of gravitational collapse, that leads to the formation of black holes. 
Even though, in black holes the singularity is hidden from an asymptotic observer by the event horizon, it signals the breakdown of the theory itself 
in very high curvature regions, a shortcoming  of GR, which is expected to be resolved in a full quantum theory of gravity. Due to the lack of a consistent theory of quantum 
gravity, one of the popular approaches in the literature is to construct phenomenological regular spacetime metrics, where  the usual 
curvature invariants are finite everywhere in spacetime. However, often the cost of these by-hand constructions is that these 
regular solutions violate at least one or all of the standard energy conditions, indicating the presence of exotic source of 
matter. For  reviews and further references of such constructions, see \cite{Nicolini:2008aj,Ansoldi:2008jw,Maeda:2021jdc,Sebastiani:2022wbz,Torres:2022twv}. 

Typically, the presence of a curvature singularity is synonymous with the fact that the corresponding causal geodesics are incomplete, 
i.e., time-like or null geodesics cannot be well defined after a finite proper time for a test observer \cite{Hawking:1970zqf}. 
However, in a recent work \cite{Zhou:2022yio}, it was shown that even some of the popular regular metrics studied widely  
in the literature, e.g., the Hayward metric \cite{Hayward:2005gi} and the Culetu-Ghosh-Simpson-Visser (CGSV) regular black hole metric with asymptotically Minkowski core \cite{Simpson:2019mud, Culetu:2013fsa, Ghosh:2014pba}, 
are not geodesically complete for negative values of the radial coordinates $r$. Consequently, the true meaning of `regular black hole' 
in this context needs to be addressed more carefully. In the same work \cite{Zhou:2022yio}, simple modifications of the above two 
metrics were also proposed, and it was shown that  it is possible to extend the casual geodesics to all the values of proper time, 
thereby making the spacetime geodesically complete.

In this paper our motivation is to explore a systematic way of deforming a geodesically incomplete spacetime into a geodesically complete spacetime, 
by using the  Simpson-Visser (SV) procedure of regularising a spacetime singularity. Consider for example the 
SV modification \cite{SV1, SV2, Franzin2, mazza, shaikh:MNRAS2021, jcch, Bronnikov:2021uta} that can be used to regularise the curvature singularity of the Schwarzschild and other
spacetime by introducing a single non zero parameter that assumes a constant value \cite{SV1}. \footnote{The SV method has gained a lot of recent attention. For a non-exhaustive list of recent works, see - \cite{Ye:2023xyv} - \cite{Junior:2022zxo}.} The resultant metric, 
depending on the value of this constant parameter, can interpolate between a wormhole branch and a regular black hole branch, and 
in all the cases the metric is geodesically complete for the full ranges of the radial coordinate. 
However, the qualitatively different CGSV black hole \cite{Simpson:2019mud}, which represents an 
exponential suppression of mass to smear out the Schwarzschild singularity at the location $r=0$ \cite{Culetu:2013fsa, Ghosh:2014pba, Simpson:2019mud}
does not enjoy this last property. \footnote{To the best of our knowledge, the phenomenological version of this metric was first proposed in  \cite{Culetu:2013fsa}, along with the rotating version in \cite{Ghosh:2014pba}. However, the full significance of this metric, from carefully chosen first principles was elucidated in \cite{Simpson:2019mud, Simpson:2021dyo, Simpson:2021zfl}.} As explained in \cite{Zhou:2022yio}, the CGSV spacetime is strictly valid only in the region of positive values of $r$. 
As a result, the effective potential encountered by  a massive particle has a singularity upon extending $r$ to negative values, in a finite proper time.  

In this paper we show that if we further modify the CGSV metric by the SV method itself, we can cure its geodesic incompleteness for $r<0$. 
The resulting metric may represent a wormhole or a regular black hole with two horizons (per side of the throat), and a single horizon (per side of the 
throat). In all the cases, the curvature invariants are shown to be regular for all values of the radial coordinates, and 
also the casual geodesics are shown to be complete everywhere.  We find out the possible nature of the source of 
this metric for which this is an exact solution of the Einstein equations. Furthermore, the motion of photons and the resulting effective potential 
encountered by them in this geometry are discussed as well.

The method proposed here should be contrasted with the one used in \cite{Zhou:2022yio}. In that paper, a somewhat different formalism
was used to make the CGSV solution geodesically complete. With a generic static, spherically symmetric solution of the form 
\begin{equation}\label{static_metric}
	ds^2=-\mathcal{A}(r)\text{d}t^2+\mathcal{A}^{-1}(r)\text{d}r^2+\mathcal{B}^2(r)\text{d}\Omega^2~,
\end{equation}
the CGSV spacetime is represented  by $\mathcal{A}(r) = 1-(2M/r)e^{-a/r}, \mathcal{B}^2(r)=r^2$, with $M$, and $a$ (not to be confused with the angular momentum of a rotating metric) are positive, real constants. As shown in \cite{Zhou:2022yio}, the modification
$\mathcal{A}(r) = 1-(2M/r)e^{-a^2/r^2}$ cure the geodesic incompleteness of the CGSV solution in regions $r<0$. 
In that paper, a different prescription was put forward to make the Hayward black hole \cite{Hayward:2005gi} 
geodesically complete. However, instead of a case by case resolution, 
here we propose that,  a more systematic way to cure geodesic incompleteness that may exist in {\it any} regular solution
of the form in Eq. \eqref{static_metric} is to 
perform the replacement $r \to \sqrt{r^2+r_{0}^2}$, with a parameter $r_{0}\geq 0$. With this replacement, now we have the modified function
$\mathcal{A}(r) = 1-(2M/\sqrt{r^2+r_{0}^2})e^{-a/\sqrt{r^2+r_{0}^2}}$, which is a two parameter family of solutions. Here, $a$ and $r_{0}$ are real and 
positive non-zero constant parameters. Indeed, the interplay between $a, r_{0}$ and $M$ leads to interesting branches 
of the geodesically complete spacetime. 
Furthermore, as we will argue, this modification renders any possible geodesically incomplete regular solutions to
geodesically complete ones as well.

\section{The modified SV metric with asymptotic Minkowski core metric}\label{metric}

Following the discussion above, we focus on the CGSV spacetime, and consider a closely related cousin - a two-parameter extension of 
the Schwazschild metric of the form
\begin{equation}\label{SV_complete}
ds^2=-\Big(1-\frac{2M e^{-\frac{a}{\sqrt{r^2+r_{0}^2}}}}{\sqrt{r^2+r_{0}^2}}\Big) \text{d}t^2 + \Big(1-\frac{2M e^{-\frac{a}{\sqrt{r^2+r_{0}^2}}}}{\sqrt{r^2+r_{0}^2}}\Big)^{-1} \text{d}r^2 
+ \Big(r^2+r_{0}^2\Big) \text{d}\Omega^2~.
\end{equation}
In the limit $r_{0} \rightarrow 0$, the above metric reduces to the 
CGSV metric introduced in \cite{Simpson:2019mud, Culetu:2013fsa, Ghosh:2014pba}. On the other hand, for $a \rightarrow 0$, this metric reduces to the SV
modification of the Schwarzschild metric first introduced in \cite{SV1}. 
Our goal is to show that this metric is geodesically complete,  
 discuss its possible extension beyond $r=0$, and find the sources for which this metric is an exact solution of the Einstein equations. In constructing this metric, we have motivated from the SV procedure of regularising a curvature singularity by introducing a wormhole throat at the location of singularity \cite{SV1}, and consequently followed the SV procedure of replacing the radial coordinate $r$ to $\sqrt{r^2+r_{0}^2}$, without changing the $\text{d}r^2$ part.

\subsection{Structure of the metric}
To better understand the structure of the metric of Eq. (\ref{SV_complete}), we first use a coordinate transformation $k^2=r^2+r_{0}^2$
to rewrite the radial part as
\begin{equation}
	g_{kk}=\Big(1-\frac{2M e^{-\frac{a}{k}}}{k}\Big)^{-1} \frac{k^2}{k^2-r_{0}^2}~~.
\end{equation}
The inverse component of the metric $g^{kk}$, has zeros at the location of $\bar{\Delta}=1-\frac{2M e^{-\frac{a}{k}}}{k}=0$, as well as  at $k=\pm r_{0}$, and therefore, the final nature of the metric is determined by the interplay of these roots.
To find out the the roots of the equation 
\begin{equation}\label{horizon}
	1-\frac{2M e^{-\frac{a}{k}}}{k}=0~, 
\end{equation}
which we label as $k_{+}$ and $k_{-}$, we will solve this equation numerically.\footnote{Note that, in the following we 
	shall  choose the parameter values in such a way that this equation has only two real roots.  } 
  This condition is the same as that of the horizon condition for the original 
CGSV metric \cite{Simpson:2019mud}, however in a different coordinate system from the one used here. Note also that for Eq. (\ref{horizon}) to have real roots, 
the condition  $a< \frac{2M}{e}$ must be satisfied, where, $e$ is the Euler's number. 
For $a>\frac{2M}{e}$, there is no real root of the Eq. (\ref{horizon}), while for $a<\frac{2M}{e}$, there exists real positive roots, which coincides with each other for $a=\frac{2M}{e}$.
Though it is also possible to write down the exact solution of Eq. (\ref{horizon}) in terms of the Lambert W functions, 
as illustrated in \cite{Simpson:2019mud}, in this paper we work only with some chosen numerical values of the parameters to
illustrate the basic properties of the spacetime  in a simple manner. 
The nature of the line element in Eq. \eqref{SV_complete}, for different branches of the parameter space can be classified as described below,  where for the ease of illustration we have assumed $M=1$. 

Now we consider the above mentioned two cases, $a<2M/e$ and $a>2M/e$ separately.
For the first case, as an illustration, we  take the value of $a=0.5 < \frac{2M}{e}$. 
In this case we see that Eq. (\ref{horizon}) has  two real, positive roots, with numerical values  $k_{-}\sim 0.232$ and $k_{+} \sim 1.399$ 
respectively.
The value of the other parameter $r_{0}$ is not fixed and depending on the relative values of $r_{0}$, $k_{-}$ and $k_{+}$, 
several different situations  might appear. 

\noindent
$\bullet$ $a< \frac{2M}{e}$, $r_{0}>k_{+}$ :
In this case the metric represents a \textit{two way} wormhole with a throat at $r=0 ~ (k=r_{0})$, as can be seen by the 
following analysis. (i) As long as the above mentioned condition is satisfied, there is no possibility of a horizon present at spacetime. 
The metric can be extended from  $r = \infty$ to $r =-\infty$, through $r=0$. (ii) The analysis of all the associated curvature 
scalars confirms that there is no curvature singularity present in the spacetime. 

\noindent
$\bullet$ $a< \frac{2M}{e}$, $k_{-}<r_{0}<k_{+}$ :
In this case the metric has a single horizon at the location of $k_{+}$, and the throat of a wormhole at the location $r=0 ~ (k=r_{0})$. 
Here, the metric can be extended from  $r = \infty$ to $r =-\infty$ through $r=0$.  However, due to the presence of the event horizon, 
the wormhole is only one way traversable. Similarly, the curvature scalar can be shown to be finite in this case also. 

\noindent
$\bullet$ $a< \frac{2M}{e}$, $r_{0}<k_{-}<k_{+}$ :
For this parameter range, the metric consists of two event horizons at the locations $k_{-}$ and $k_{+}$. 
The metric also has a timelike throat at the location $r=0$, and the range of the radial coordinate is $-\infty<r<\infty$, thus the metric is a one way traversable wormhole. 

\noindent $\bullet$
Finally, we consider the case
 $a> \frac{2M}{e}$ :
In this case, which is qualitatively different from the others, the metric does not have a horizon for any values of 
the other parameters, and the final metric always represents a wormhole spacetime. 

It is important to note that the metric reduces to the Schwarzschild solution for large values of the radial coordinate values $r$. However, in contrast to that of the construction \cite{Simpson:2019mud, Zhou:2022yio}, the metric in this case, in general, does not have a Minkowski core. The global structure of this spacetime can be visualized by drawing the corresponding Penrose diagrams. For different branches  of the spacetime (wormhole, regular black holes with single
or double horizons per side of the throat)
discussed above, the Penrose diagrams are qualitatively similar to the black bounce and the charged black-bounce solution discussed in \cite{SV1, Franzin2}, and we do not repeat them here.

\subsection{ Motion of  a massive particle and geodesic completeness}
To understand geodesic completeness, and hence the singularities of our proposed line element in Eq. \eqref{SV_complete}, we now consider the 
radial motion of a  massive test particle moving freely in this spacetime.
For illustration, it is useful to consider the general form of a static spherically symmetric line element 
of Eq. (\ref{static_metric}). 
For our case the explicit expressions for the unknown functions $\mathcal{A}(r)$ and $\mathcal{B}(r)$ can be read off directly.
 
 The equation of motion of a radially moving massive free 
particle in this geometry can be written as (an overdot denotes derivative with respect to the proper time $\tau$)
\begin{equation}
	\dot{r}^2=E^2-\mathcal{A}(r)~,
\end{equation}
where $E=\mathcal{A}(r)\dot{t}$ is the conserved specific energy along the particle trajectory. From this expression, we see that the 
effective potential encountered by the massive particle is $V_{eff}(r)=\mathcal{A}(r)$. Thus, for the line 
element in  Eq. (\ref{SV_complete}), we have the effective potential to be
\begin{equation}
	V_{eff}(r)=1-\frac{2M e^{-\frac{a}{\sqrt{r^2+r_{0}^2}}}}{\sqrt{r^2+r_{0}^2}}~.
\end{equation}
The dependence  of the effective potential on  the coordinate $r$ (with possible extension to negative values of $r$)
determine the  behaviour of the motion of the  massive particle in this geometry, and hence whether the metric is  geodesically complete.

The first point to note is that the function $\mathcal{A}(r)$, and hence the effective potential, is continuous across $r=0$. 
Specifically, $V_0=V_{eff}\big|_{r \rightarrow 0}=1-\frac{2 M}{r_{0}}e^{-a/r_{0}}~.$ This fact is in contrast with the CGSV metric, 
which is just the $r_{0} \rightarrow 0$ limit of the metric in Eq. (\ref{SV_complete}). In fact, it can be easily checked that the 
effective potential of CGSV spacetime  is divergent at $r \rightarrow 0^{-}$. 

Next, we find out the proper time a radially moving massive particle takes to reach $r=0$. Assuming that the particle starts
from an initial location of $r_i$, the proper time  it takes to reach a final location $r$ is given by the formula
\begin{equation}\label{proper_time}
\tau(r)=\int_{r_i}^{r}\frac{dr}{\sqrt{E-V_{eff}(r)}}~.
\end{equation}
Depending on the sign of $V_0$ it is useful to consider three different cases. 

\noindent
$\bullet$ $V_0>0$ : Since at asymptotic infinity, the metric of Eq. (\ref{SV_complete}) reduces to the Minkowski metric, 
the effective potential at $r=0$ can be positive when, for some particular choices of parameters, $a,r_{0},M$,
the function $\mathcal{A}(r)$ has no real root or has two roots. Thus in this case, the spacetime is either horizonless or has an 
even number of horizons. One such situation is illustrated in Fig. \ref{SV_complete} by the red 
curve. The choice of the parameters are made in such a way that the metric has two horizons for positive values of $r$.
We can also choose these parameters in such a way that there are no real roots of the function $\mathcal{A}(r)$, so that in that 
case as well, the effective potential at $r=0$ is positive.

\begin{figure}[h!]
	\begin{minipage}[b]{0.45\linewidth}
		\centering
		\includegraphics[width=0.9\textwidth]{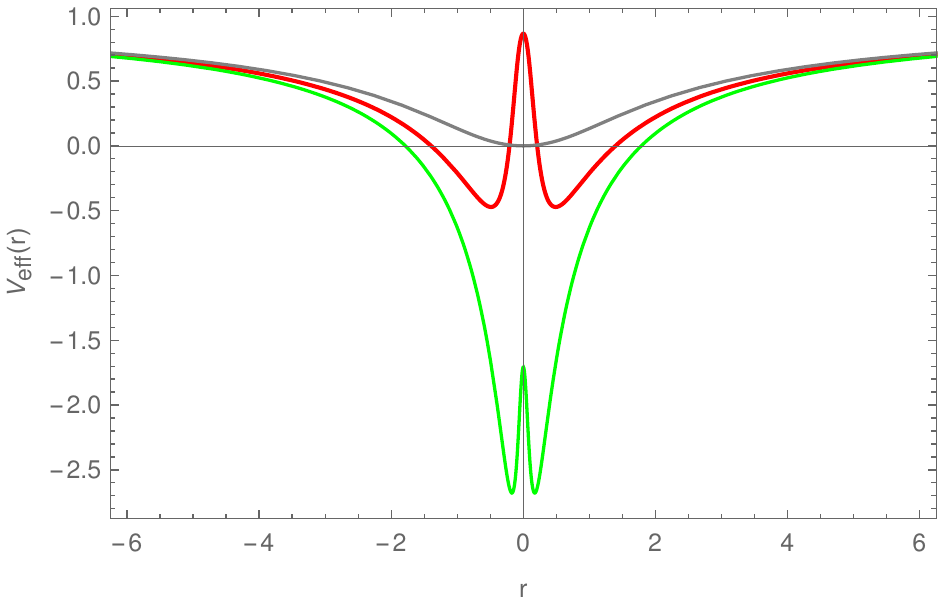}
		\caption{Plot of the effective potential $V_{eff}$. The choices of the parameters 
			for three different curves shown are : $a=0.5, r_{0}=0.1, M=1$ (Red), and
		$a=0.693, r_{0}=1, M=1$ (Gray), and $a=0.2, r_{0}=0.1, M=1$ (Green).  }
		\label{fig:V0_PLOTS}
	\end{minipage}
	\hspace{0.2cm}
	\begin{minipage}[b]{0.45\linewidth}
		\centering
		\includegraphics[width=0.9\textwidth]{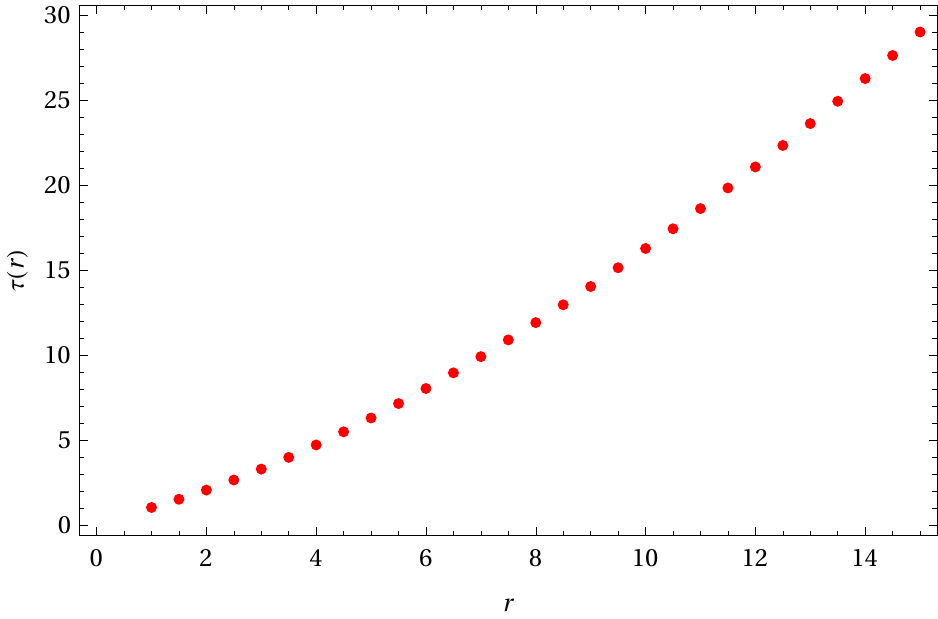}
		\caption{The plot of the proper time a radial test particle takes to reach $r=0$ starting from different initial positions $r$. 
			Here, the parameter values  are the same as the red curve in Fig. \ref{fig:V0_PLOTS}, and $E(=1)>V_{eff}\big|_{r=0}$.}
		\label{fig:propertime}
	\end{minipage}
\end{figure}

To study the particle motion in this case we first consider that the energy $E$ of the particle is less than $V_0$,
and it starts from $r_i>0$. Here the test
particle cannot reach at $r=0$, and will bounce back before crossing either of the two horizons, or after crossing both the horizons. 
When $E=V_0$, the particle takes an infinite proper time to reach $r=0$, as seen from Eq. (\ref{proper_time}). Finally, 
when $E>V_0$ the particle takes a finite amount of proper time to reach at $r=0$ from the
initial position. In Fig. \ref{fig:propertime} we have plotted the proper times the particle takes to reach at $r=0$
starting from different initial positions, by numerically integrating Eq. (\ref{proper_time}). The proper time is finite in all cases. 

Since the  metric function $\mathcal{A}(r)$ is continuous across $r=0$ and the effective potential is finite everywhere, 
we can extend the metric across $r=0$ to negative 
values of $r$ as well.  Thus a test particle can move from $-\infty$ to $\infty$ in this spacetime, and this takes an infinite 
amount of proper time.

\noindent
$\bullet$ $V_0<0$ : Next we consider the case when the effective potential is negative at $r=0$ (the green curve in Fig. \ref{fig:V0_PLOTS}). In this case,  
the  parameters take values in such a way that the metric function $\mathcal{A}(r)$ has an odd number of real zeros on the positive $r$ axis. 
From an analysis similar to the previous case, we see that a test particle takes 
a finite amount of proper time to reach $r=0$, and hence we can extend the spacetime beyond $r=0$. 

\noindent
$\bullet$ $V_0=0$ : Finally, when the effective potential vanishes at $r=0$, the metric has a single horizon at that location.
This case is shown in Fig. \ref{fig:V0_PLOTS}  with the grey curve. 
The metric is continuous across $r=0$, and can be extended to negative values of $r$.

From the above analysis we see that the metric in Eq. (\ref{SV_complete}), unlike the original CGSV metric, is geodesically 
complete, and has an extension beyond $r=0$ to negative values of $r$. 
 Here, we take the extension to the negative $r$ axis to be the same as in Eq. (\ref{SV_complete}).  
Furthermore, unlike the geodesically complete extension of CGSV metric proposed in \cite{Zhou:2022yio}, our metric is symmetric 
about $r=0$, so that an observer must cross an even number of event horizons (if they are present) to reach $r=\infty$ to $-\infty$. 
It is of course possible to make other geodesically complete extensions of our metric, which are not symmetric about $r=0$.

\subsection{Energy momentum tensor and the source of the geodesically complete metric}

In this subsection we compute the energy-momentum tensor (EMT) corresponding to the line element in Eq. (\ref{SV_complete}), and
discuss its interpretation in terms of possible sources of such a solution of the Einstein equations.

If we assume that our proposed metric satisfies the Einstein equation with the energy momentum tensor $T_{\mu \nu}$, 
then in the general case assuming that the spacetime under consideration has two horizon locations on the positive $r$ axis (at $r_{+}$ and $r_{-}$ respectively),
\footnote{Since the metric extension is symmetric about $r=0$ there are two horizons for negative values of $r$, at $-r_{+}$ and $-r_{-}$ respectively, as well.} we 
can obtain the  following relation 
\begin{equation}\label{rho+pr}
G^{t}_{t}-G^{r}_{r}=2\mathcal{A}(r)\frac{\mathcal{B}^{\prime\prime}(r)}{\mathcal{B}(r)}=\mp(\rho+p_r)~,
\end{equation} 
where $\rho$ is the energy density and $p_r$ is the radial pressure, and a prime denotes a derivative with respect to the radial coordinate. 
Furthermore, in the above equation  the minus sign is valid when we are considering the region outside the outer horizon $r_{+}$, or the 
region between $r_{-}$ and $-{r_{-}}$ or region outside $r_{-}$ (from $r_{-}$ to $r=-\infty$). These are precisely the regions where 
the metric function $\mathcal{A}(r)$ is positive. In the other two regions, where $\mathcal{A}(r)$ is negative, the plus sign has to be used. 
Hence, it is useful to write  the above  equation in the following suggestive form
\begin{equation}\label{rho+pr_mod}
2|\mathcal{A}(r)|\frac{\mathcal{B}^{\prime\prime}(r)}{\mathcal{B}(r)}=-(\rho+p_r)~,
\end{equation}
In this form, it is easy to see that though the individual components of the EM tensor are determined by the sign of $\mathcal{A}(r)$, the
sign of the sum of radial pressure and energy density does not depend on it. Rather its sign  is fixed only by the sign 
of the second derivative of the areal radius. As we shall see below, this is also related to the nature of the source of the metric.

In our case the quantity $\frac{\mathcal{B}^{\prime\prime}(r)}{\mathcal{B}(r)}$ is everywhere positive, and hence the NEC  is violated 
everywhere except the horizons (where it is satisfied marginally) for the metric considered in this paper. Furthermore, since for a 
geodesically complete black hole, the effective potential $\mathcal{A}(r)$ encountered by a radially moving test particle must be 
finite and continuous, it can be seen that the quantity $\rho+p_r$ is smooth everywhere. Note that the opposite is also true, i.e., 
for a geodesically incomplete black hole, like the CGSV metric, $\rho+p_r$  is usually discontinuous and divergent for some values 
of the radial coordinate $r$. For the CGSV metric the discontinuity is at $r=0$.  

These conclusions can be verified by explicitly computing the components of the EMT. 
Consider the case when the spacetime has two sets of inner and outer horizons. Assuming that the metric satisfies the 
Einstein equations $G^{\nu}_\mu=T^{\nu}_\mu$, for regions outside the outer horizon $r_{+}$, or  between $r_{-}$ and $-{r_{-}}$ or  
outside $-r_{-}$ (from $-r_{-}$ to $r=-\infty$), the energy density and the radial component of the pressure are given by
\begin{equation}
	\rho = \frac{1}{(r^2+r_{0}^2)^2}\bigg[\frac{2aM r^2}{r^2+r_{0}^2}e^{-\frac{a}{\sqrt{r^2+r_{0}^2}}}-r_{0}^2\bigg(1-\frac{4Me^{-\frac{a}{\sqrt{r^2+r_{0}^2}}}}{\sqrt{r^2+r_{0}^2}}\bigg)\bigg]~,~
	p_r=-\frac{r_{0}^4+r_{0}^2 r^2 +2aM r^2 e^{-\frac{a}{\sqrt{r^2+r_{0}^2}}}}{(r^2+r_{0}^2)^3}~.
\end{equation}
From these we obtain 
\begin{equation}
	\rho+p_r=-\frac{2r_{0}^2}{(r^2+r_{0}^2)^2}\bigg[1-\frac{2M e^{-\frac{a}{\sqrt{r^2+r_{0}^2}}}}{\sqrt{r^2+r_{0}^2}}\bigg]~.
\end{equation}
It is easy to check that this expression is in accord with Eq. (\ref{rho+pr}).
Furthermore, using this expression, we see that, except at the locations of the inner and outer horizons, in the regions mentioned above, 
the quantity $\rho+p_r$ is always negative. Thus, except at the horizon locations, in the three regions mentioned above, the  
NEC is everywhere violated. In the region between two horizons, magnitudes of $\rho$ and $p_r$ get interchanged, and in that case also, 
it can be explicitly checked that NEC is everywhere violated.
 
We can now find out the source of the metric given in Eq. (\ref{SV_complete}). Following, \cite{Bronnikov:2022bud}, where it was shown that a 
scalar and a nonlinear electromagnetic field can act as the source of an arbitrary static spherically symmetric metric of the general form given in 
Eq. (\ref{static_metric}), we consider the following matter action
\begin{equation}
	\mathcal{S}_m=\int\sqrt{-g}d^4x\bigg[-\frac{1}{2}h(\phi)\partial_\mu\phi\partial^\mu\phi-V(\phi)
	+\mathcal{L}(\mathcal{F})\bigg]~.
\end{equation}
Here,  $h(\phi)$ and $V(\phi)$ are functions of the scalar filed $\phi$, and $\mathcal{L}(\mathcal{F})$ is the Lagrangian density 
of the nonlinear electromagnetic field. The function $h(\phi)$ determines the nature of the scalar field (which is self interacting and minimally 
coupled to gravity), and $V(\phi)$ is the potential term in the scalar field Lagrangian density. The scalar field is a function of the 
radial coordinate, and $\mathcal{L}(\mathcal{F})$ is a function of $\mathcal{F}=F_{\mu\nu}F^{\mu\nu}$, $F_{\mu\nu}$ being the 
electromagnetic field tensor - defined in terms of the vector potential as $F_{\mu\nu}=\partial_\mu A_{\nu}-\partial_\nu A_{\mu}$.

Performing a standard analysis \cite{Bronnikov:2022bud}, we obtain the following two relations
\begin{eqnarray}\label{Einstein_eqs1}
G^{t}_t-G^{r}_r=T^{t}_t-T^{r}_r~ \rightarrow ~ \frac{\mathcal{B^{\prime\prime}}(r)}{\mathcal{B}(r)}
=-\frac{1}{4}h(r)\phi^{\prime 2}(r)~,\\
~~\text{and}~~ G^{t}_t-G^{\theta}_\theta=T^{t}_t-T^{\theta}_\theta=-\mathcal{F}\mathcal{L}_{\mathcal{F}}~.
\label{Einstein_eqs2}	
\end{eqnarray}
Utilising the parameterisation freedom of the scalar field, in the following, we take  the customary choice of the 
	scalar field $\phi(r)=\arctan (r/c)$. Furthermore, as the source of the nonlinear electromagnetic field, we consider
	a magnetic monopole of charge $q_m$, so that the invariant is given by $\mathcal{F}=2q_m^2 /\mathcal{B}^4(r)$. Given the metric 
	components, $\mathcal{A}(r)$ and $\mathcal{B}(r)$, the general procedure of obtaining the unknown function appearing in the 
	Lagrangian has been discussed in \cite{Bronnikov:2022bud}. E.g., given  functional 
	dependence of $\phi(r)$, as above, the function $h(r)$ obtained from Eq. \eqref{Einstein_eqs1}
	is given by $h(r)=-\frac{4}{\phi^{\prime 2}(r)} \frac{\mathcal{B^{\prime\prime}}(r)}{\mathcal{B}(r)}$,
	while, with $\mathcal{F}=2q_m^2 /\mathcal{B}^4(r)$, the function $\mathcal{L}_{\mathcal{F}}$ can be obtained 
	by solving the differential equation $\mathcal{L}_{\mathcal{F}}=-\frac{1}{\mathcal{F}}(T^{t}_t-T^{\theta}_\theta)$. The remaining function
	$V(\phi)$ can  be determined by using any of the  components of the Einstein equations.
	 Below we discuss the  nature of these functions for the line element 
	Eq. \eqref{SV_complete}.

\begin{figure}
		\begin{minipage}[b]{0.45\linewidth}
		\centering
	\includegraphics[width=0.9\textwidth]{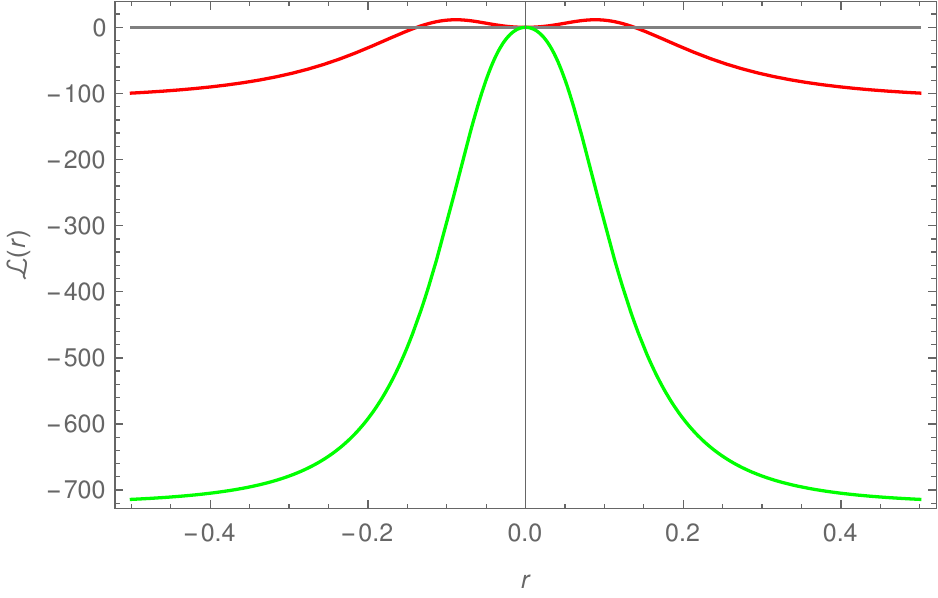}
	\caption{Plot of the NED Lagrangian. The parameters corresponding to the red and green curves are same as the  red and green curves 
		in Fig. \ref{fig:V0_PLOTS}.  }
	\label{fig:ned}	
		\end{minipage}
  \hspace{0.2cm}
   \begin{minipage}[b]{0.45\linewidth}
   	\centering
   		\includegraphics[width=0.85\textwidth]{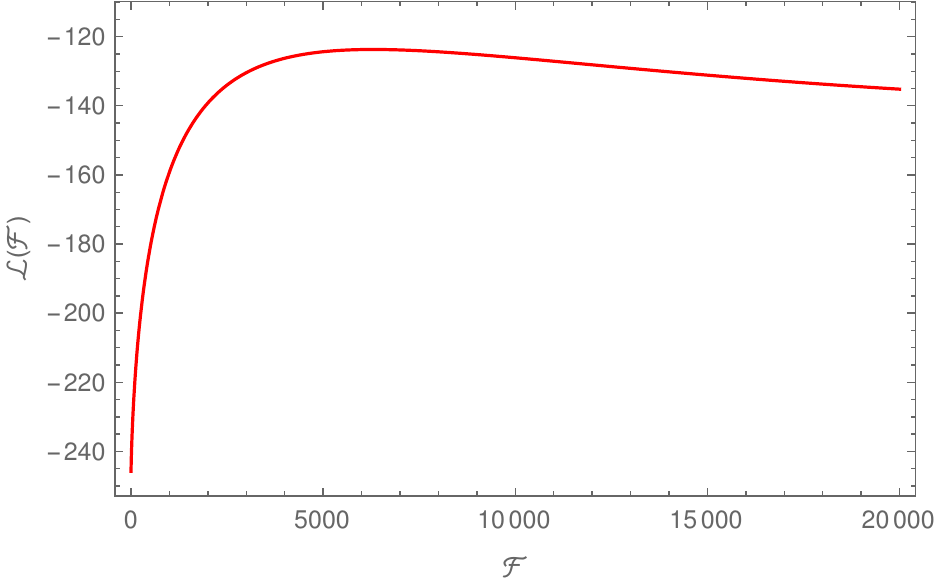}
   	\caption{Plot of the NED Lagrangian $\mathcal{L}(\mathcal{F})$ as a function of $\mathcal{F}$. Here, $q_m=1$, and other parameters are the  same as the  red curve 
   		in Fig. \ref{fig:V0_PLOTS}. We also see that, here, $\mathcal{L}(\mathcal{F} \rightarrow 0)=-245.8$.}
   	\label{fig:nedF}
   		\end{minipage}
\end{figure}

 From  Eq. (\ref{Einstein_eqs1}) it is easy to see that $h(r)$ in this case is just a 
negative constant.\footnote{Notice that though sign of $\rho+p_r$ depends on the metric function $\mathcal{A}(r)$, once we have made the above 
choice of the scalar field, the sign of the function $h(r)$ is determined only by the areal radius $\mathcal{B}(r)$ and its second derivative.} 
Thus, the scalar field is always phantom in nature. On the other hand, from  Eq. (\ref{Einstein_eqs2}) we can obtain the
NED Lagrangian as the solution of the following differential equation
\begin{equation}\label{dif_ned}
	\frac{\partial \mathcal{L}(r)}{\partial r}=\frac{1}{(r_{0}^2+r^2)^{9/2}}\Big[4 M r e^{-\frac{a}{\sqrt{r^2+r_{0}^2}}}
	\Big(ar^2\big(a-4\sqrt{r^2+r_{0}^2}\big)+c^2\big(a\sqrt{r^2+r_{0}^2}-3r^2\big)-3r_{0}^4\Big)\Big]~.
\end{equation}
We can write the analytical solution of this equation in terms of incomplete Gamma functions as
\begin{equation}\label{NEDdensity}
  \begin{split}
  		\mathcal{L}(r)=-4M \bigg[\frac{4e^{-f(r)}}{(r_{0}^2+r^2)^{3/2}}+
  		\frac{12 e^{-f(r)} 
  		\Big(a^2+2a\sqrt{r_{0}^2+r^2}+2(r_{0}^2+r^2)\Big)}{a^3(r_{0}^2+r^2)}\\
  	-\frac{1}{a^5}\Big[(a^2-3r_{0}^2)\Gamma(5, f(r))+5r_{0}^2\Gamma(6, f(r))-r_{0}^2 \Gamma(7, f(r))\Big]\bigg]+\mathcal{C}~,
  \end{split}
\end{equation}
where $\mathcal{C}$ is an integration constant, and we have denoted $f(r)=\frac{a}{\sqrt{r^2+r_{0}^2}}$.
In Fig. \ref{fig:ned}, we have plotted the numerical solution of the differential equation 
	Eq. (\ref{dif_ned}), with  the integration constant being fixed by the condition $\mathcal{L}(r=0)=0$, for the parameter 
	sets corresponding to the single and double horizon (on the positive $r$ axis) cases  respectively, from Fig. \ref{fig:V0_PLOTS}. 
	It is easy to see that, for  both of these sets of parameter  values, the solutions for the  NED field are analytic functions 
	of the radial coordinate and have negative values at the limit of $r \pm \infty$. For the parameter set, where the spacetime has 
	two horizon locations on the positive $r$ axis, $\mathcal{L}(r)$ is positive around $r=0$, whereas, when there is only one horizon 
	location at the $r$ axis, the NED Lagrangian is always negative for the boundary condition we have specified in the numerical solution.

Furthermore, it is interesting  study the functional dependence of the Lagrangian density as a function of the EM invariant $\mathcal{F}$. To obtain this, we need to invert the relation $\mathcal{F}=\frac{2q_m^2}{(r^2+r_{0}^2)^2}$, and substitute in the exact expression in Eq. \eqref{NEDdensity}. The resulting 
analytical expression for $\mathcal{L}(\mathcal{F})$  is too complicated to provide here, rather we study the function numerically by plotting it in Fig. \ref{fig:nedF}, for parameter values such that the metric has double horizons in the positive $r$ axis.
	It is easy to see that for all physically admissible values of $\mathcal{F}$, 
	with the parameter values indicated in the caption,  $\mathcal{L}(\mathcal{F})$  is a regular function
	of $\mathcal{F}$, and takes only negative values. In particular in the limit $\mathcal{F} \rightarrow 0$, for the numerical values used in this 
	plot we have $\mathcal{L}(\mathcal{F} \rightarrow 0)=-245.8$.
	Notice also that for all values of the radial coordinate, with $q_m>0$, $\mathcal{F}$ is always  positive, and in the limit $r 
	\rightarrow \pm \infty $, it goes to zero. Furthermore, there is a maximum value of $\mathcal{F}$ which it takes  in the limit $r 
	\rightarrow 0$. We see that within this range of $\mathcal{F}$, the Lagrangian density of the NED field is an analytic function. 
	These conclusions are true for other branches of the  spacetime under consideration as well.

\section{Photon motion in the geodesically complete background}

In this section, we discuss the motion of a photon in the background of the metric in Eq. (\ref{SV_complete}),  which is essential to study the shadow 
structure of the regular and geodesically complete spacetime we have introduced in this paper. It is also important to compare and contrast the null 
geodesics for different branches of spacetime, in particular, the effect of geodesic completion by implementing a wormhole throat. Using the standard method of 
finding null geodesics for a generic spherically symmetric wormhole spacetime \cite{shaikh3}, we can write down the effective potential for an 
asymptotically flat metric of the form in Eq. (\ref{static_metric}) in the units of angular momentum $L$ of the particle  as
\begin{equation}\label{effectivepotential}
V_{p}(r)=L^2 \frac{\mathcal{A}(r)}{\mathcal{B}^2(r)}~.
\end{equation}
For the case of metric represented by Eq. (\ref{SV_complete}), the expression of $V_{p}(r)$ is given by 
\begin{equation}\label{effectivepotentialgc}
V_{p}(r)=\frac{L^2 }{r^2+r_{0}^2}\Big(1-\frac{2M }{\sqrt{r^2+r_{0}^2}}e^{-\frac{a}{\sqrt{r^2+r_{0}^2}}}\Big)~.
\end{equation}
Now depending on the range of the parameters, three different cases may appear, which we describe below.

\begin{figure}[h!]
	\begin{minipage}[b]{0.45\linewidth}
		\centering
		\includegraphics[width=0.75\textwidth]{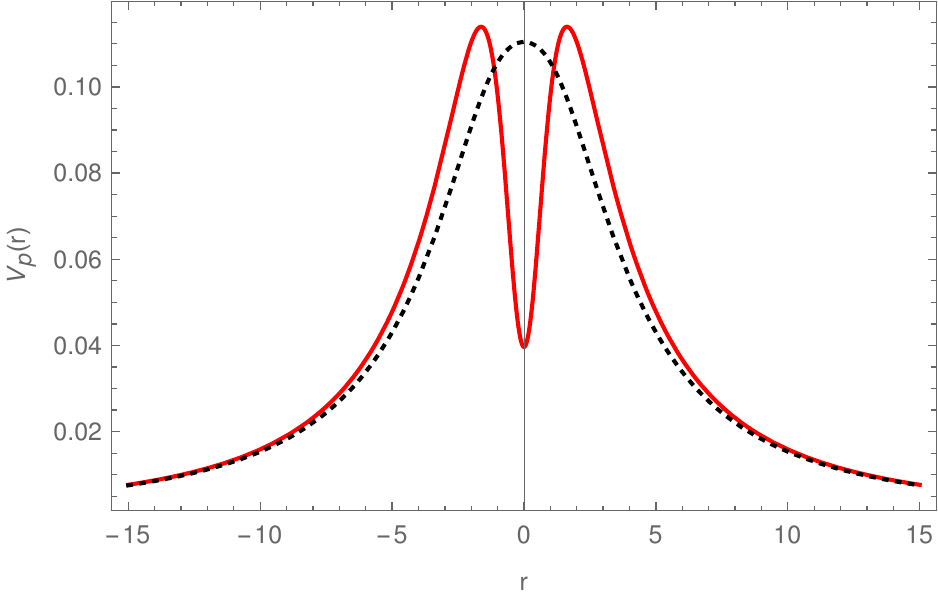}
		\caption{Effective potential $V_{p}$ in the wormhole branch. The parameters are : $r_{0}=1.5$ (red), $r_{0}=2.5$ (black), $a=0.5, M=1$, and  $L=1$.  }
		\label{fig:Veff_WORMHOLE}
	\end{minipage}
	\hspace{0.2cm}
	\begin{minipage}[b]{0.45\linewidth}
		\centering
		\includegraphics[width=0.75\textwidth]{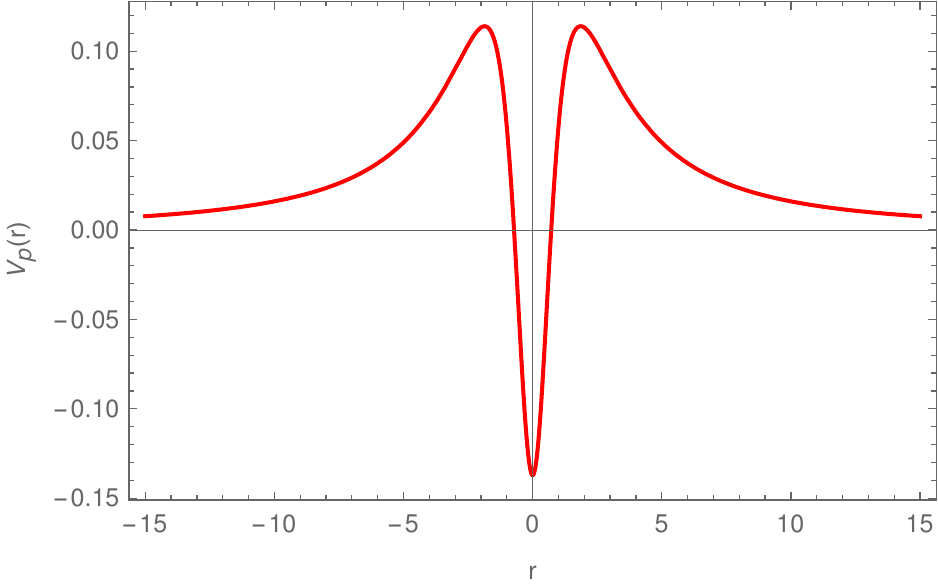}
		\caption{Effective potential $V_{p}$ when there is a single horizon per side of the throat. Parameters: $r_{0}=1.2$, $a=0.5, M=1$, and
			$ L=1$. }
		\label{fig:veff_singlehorizon}
	\end{minipage}
\end{figure}

\noindent
$\bullet$ For the wormhole branch of the full solution, we note that in general two different types of effective potentials may appear. 
As depicted in Fig. \ref{fig:Veff_WORMHOLE}, the first type includes symmetrically placed single maxima per side of the throat, and the 
throat is the location of the minima of the effective potential (shown by the red curve). Here, these locations outside the throat can act 
as the photon sphere for light coming from a source that is on the same side of the throat as that of an observer.  A light ray that has an 
impact parameter $b (= L/E)$ greater than that of a critical impact parameter ($b_{m}$) will turn at the photon sphere and will reach an 
asymptotic observer \cite{shaikh3}. The corresponding framework for strong lensing to calculate the deflection angle, as presented in \cite{shaikh3} 
will also be applicable here. On the other hand, when the source and the observer are on different sides of the throat, then photons with an impact 
parameter $b>b_{m}$ will be turned back on the same side of the throat, and will not reach the observer who is on the other side of the throat. 
However, for $b<b_{m}$, the photons will cross both the photon sphere and the throat and
reach the observer on the other side.  

Further, when the maxima of the potential coincides with the location of the throat (see the black dotted curve in Fig. \ref{fig:Veff_WORMHOLE}), the throat 
can itself act as a photon sphere \cite{shaikh2}. Then photons from a source that is on the same side of the observer,  with impact parameter $b$ greater 
than that of the critical impact parameter for the throat $b_{th}$ will be strongly lensed to the asymptotic observer. On the other hand, when the source
and the observer are on two different sides of the throat, then rays with $b>b_{th}$, cannot reach the observer on the other side. 
However, $b<b_{th}$ does not have a turning point and will reach the observer on the other side of the throat. 

\noindent
$\bullet$ When the metric represents a black hole with a single horizon per side of the throat, the corresponding effective potential for 
photon motion is shown in Fig. \ref{fig:veff_singlehorizon}. In this case, there are symmetrically placed locations of maxima per side of the 
throat, and the effective potential is minimum at the throat. Similarly as discussed above, these locations outside the throat will act as  
photon spheres for light with an impact parameter greater than the critical impact parameter, and will escape to the faraway observer. 
However, due to the presence of a horizon outside the throat, the second case as mentioned above (throat as a maximum of effective 
potential) can never appear here. 

\begin{figure}
	\centering
	\includegraphics[width=0.4\textwidth]{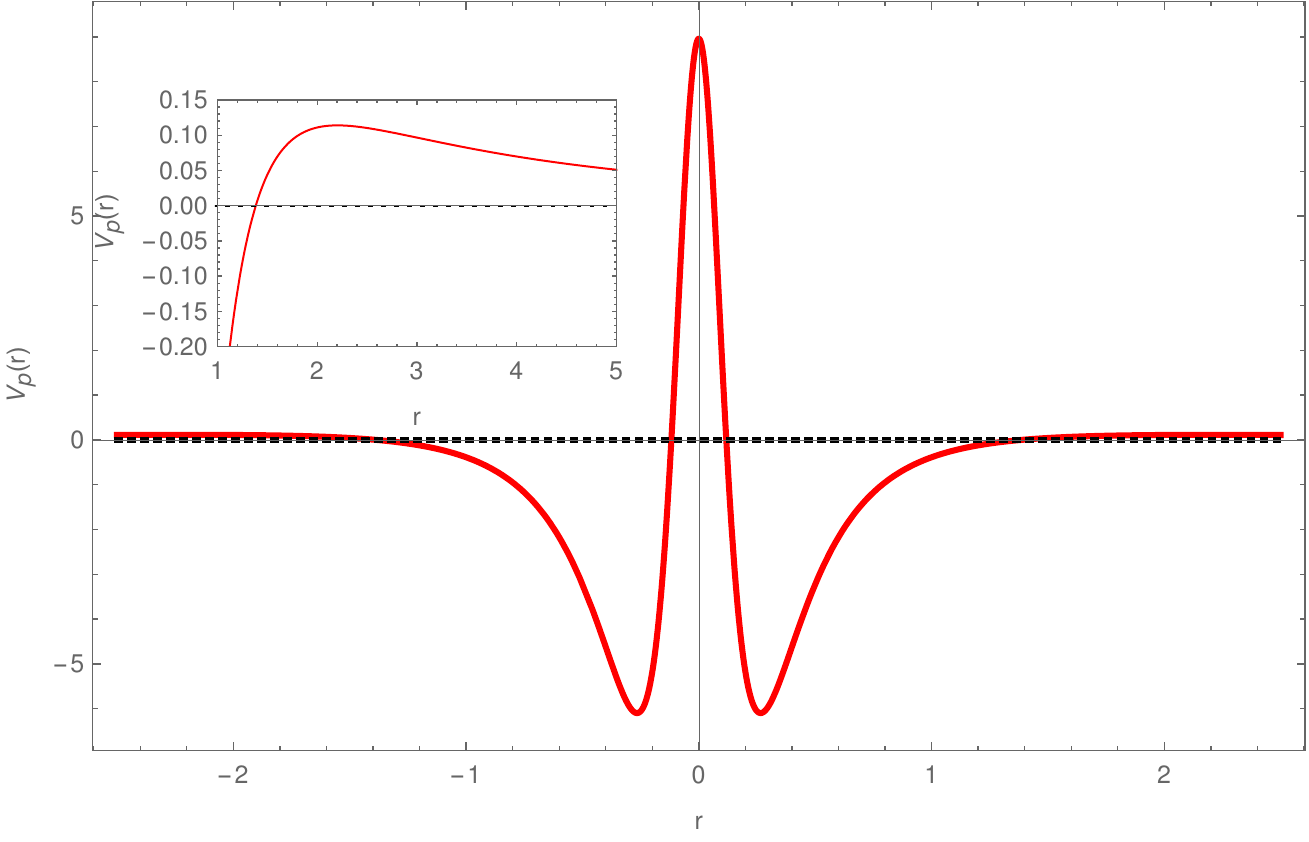}
	\caption{Effective potential when there is are two horizons per side of the throat. Parameters: $r_{0}=0.2$, $a=0.5, M=1, L=1$. The maximum of 
	the effective potential outside the outer horizon is shown in the inset. }
	\label{fig:veff_twohorizons}
\end{figure}

\noindent
$\bullet$ In the case of a spacetime that has two horizons per side of the throat, there will be one photon sphere per side of the throat, 
see Fig. \ref{fig:veff_twohorizons} inset. However, due to the presence of the event horizons on both sides of the throat, the throat cannot 
act as an effective photon sphere for the motion of null rays.  

\section{Conclusions and discussions}
In this paper, we have introduced a systematic way to construct geodesically complete metrics by following the SV 
method of regularising spacetime singularities. As an example, we have illustrated the case for the CGSV metric with an 
asymptotically Minkowski core. The resultant extension of the metric to negative values of the radial coordinate is symmetric about $r=0$, and 
the metric components are continuous through $r=0$. The spacetime also has rather interesting structure in the allowed ranges of the 
parameter space, and can interpolate between a wormhole, a black hole with a single horizon, and a black hole with two horizons per side of the 
throat present at $r=0$. All the three branches are complete in terms of the time-like geodesics, as we have shown explicitly.  We also 
find out the possible source for which the proposed geodesically complete  metric  is an exact solution of the Einstein equations.
The energy momentum tensor of the metric can be interpreted in terms of a self-interacting scalar field which minimally coupled to the 
gravity, and a NED field. For all the branches of this metric,  the motion of photons in this background is discussed in details.

Here we have illustrated the use of the SV regularisation method to construct a geodesically complete black hole using the 
CGSV metric as an example.  It will be interesting to consider extensions of other
such ``regular metrics''  and use the SV regularisation to produce geodesically complete extensions from them. 
In this context, we note that the SV method of constructing geodesically complete
extensions of conventional regular black hole spacetimes used in this
paper can be used to find out geodesically complete extensions (to negative values of the radial
coordinate) of any spherically symmetric static metric of the general form
of Eq. (\ref{static_metric}). Moreover, these extensions are symmetric
with respect to the radial coordinate, which ensures that the same form of
the metric functions can be used for smooth matching through the negative
values of the radial coordinate as well. We therefore expect the method put forward here
to be useful in a broad context.

\section*{Acknowledgements}
We thank Bidyut Dey and Rajibul Shaikh for discussions and help, and Tian Zhou for a useful correspondence. 


\end{document}